\documentclass[aps,prd,12point,nofootinbib,showpacs,superscriptaddress]{revtex4-1}
\def\theequation{\arabic{section}.\arabic{equation}}
\usepackage{psfrag}
\usepackage{subfigure}
\usepackage{color}
\usepackage{mathrsfs}
\usepackage{graphicx}
\usepackage{amssymb, bm}
\usepackage{amsmath, amsthm}
\usepackage{epstopdf}
\usepackage{hyperref}
\usepackage{enumerate}
\usepackage{longtable}
\usepackage{amsmath}

\usepackage{amsmath}  
\usepackage{amsfonts} 
\usepackage{graphicx} 
\newcommand{\R}{{\mathbb R}}

\newcommand{\N}{{\mathbb N}}

\newcommand{\be}{\begin{equation}} 
\newcommand{\ee}{\end{equation}}

\begin{document}
\def\theequation{\arabic{section}.\arabic{equation}} 

\title{Generalized Fibonacci numbers, cosmological analogies, and an 
invariant}


\author{Valerio Faraoni}
\affiliation{Department of Physics \& Astronomy, Bishop's University, 2600 
College Street, Sherbrooke, Qu\'ebec, Canada J1M~1Z7
}

\author{Farah Atieh}
\affiliation{Department of Physics \& Astronomy, Bishop's University, 2600 
College Street, Sherbrooke, Qu\'ebec, Canada J1M~1Z7
}



\bigskip
\bigskip
\begin{abstract}

Continuous generalizations of the Fibonacci sequence satisfy ODEs that are 
formal analogues of the Friedmann equation describing spatially homogeneous 
and isotropic cosmology in general relativity. These analogies are 
presented, together with their Lagrangian and Hamiltonian formulations and 
with an invariant of the Fibonacci sequence. 

\end{abstract}


\maketitle

\section{Introduction}
\setcounter{equation}{0}
\label{sec:1}

Fibonacci, also known as Leonardo Pisano or Leonardo Bonacci, introduced 
Hindu-Arabic numerals to Europe with his book {\em Liber Abaci} in 1202 
\cite{LiberAbaci}. 
He posed and solved a well-known problem involving the growth of a 
population of rabbits in idealized situations. The solution, now known  
as the Fibonacci sequence $0, 1, 1, 2, 3, 5, 8, 13,$~... is 
written  as
\begin{equation}
F_n = F_{n-1} + F_{n-2} \,, \label{1.1}
\end{equation} 
where $F_n$ is the $n$-th Fibonacci number and $F_0=0, F_1=1$. Furthermore,  
the ratio of two consecutive terms $F_{n+1}/F_n$ approaches the golden ratio 
\begin{equation}
 \varphi \equiv \frac{1 + \sqrt{5}}{2} \approx 1.61803398\, ... 
\end{equation}
as  $n \rightarrow +\infty$. 

The Fibonacci sequence can be generalized to the continuum using Binet's formula
\be
F_n =\frac{\varphi^n-\left(-\varphi\right)^{-n}}{\sqrt{5}} \,.
\ee
Furthermore, the analytic function
\begin{equation}
F_{(e)}(x)  = \frac{\varphi^x - \varphi^{-x}}{\sqrt{5}} = \frac{2}{\sqrt{5}} 
\, \sinh \left( x \ln\varphi \right) \label{Fe}
\end{equation}
reproduces part of the Fibonacci numbers, $F_n=F_{(e)}(n)$  for even $x=n \in 
\mathbb{N}$, while
\begin{equation}
F_{(o)}(x) = \frac{\varphi^x + \varphi^{-x}}{\sqrt{5}} = \frac{2}{\sqrt{5}}  
\cosh \left( x \ln\varphi \right) \label{Fo}
\end{equation} 
reproduces the other Fibonacci numbers for odd $x=n \in \N$. The function
\be
F(x) = \frac{ \varphi^x -\cos \left( \pi x \right) \varphi^{-x} 
}{\sqrt{5}} 
\ee
reproduces the entire Fibonacci sequence for  $x=n \in \N$. Here we 
focus on $F_{(e,o)}(x)$, which admit analogies with relativistic cosmology, while 
no such analogy exists for $F(x)$. 

The functions $F_{(e,o)}$ satisfy  the dual relations (where a prime denotes 
differentiation with respect to $x$)
\begin{eqnarray}
F_{(e)}'(x) &=& \left( \ln \varphi \right) F_{(o)}(x) \,, \label{ODE1}\\
&&\nonumber\\
F_{(o)}'(x) &=& \left( \ln \varphi \right) F_{(e)}(x) \,,\label{ODE2}
\end{eqnarray}
and the second order ODE
\be
F_{(e,o)}''-  \left( \ln^2\varphi \right) F_{(e,o)}=0 \,,\label{2ndorder}
\ee
of which $F_{(e)}$ and $F_{(o)}$ are two linearly independent solutions. In 
physics,  this equation 
describes the one-dimensional motion of  
a particle of position $F$ in the inverted harmonic oscillator potential 
$V(F)=-kF^2/2$ (with $K=2 \ln^2\varphi$), which is used as an example of 
an unstable mechanical system   
(this property corresponds to the fact that the Fibonacci  numbers $F_n$ 
increase without bound as $n\rightarrow \infty$). 

We will also use
\be
f(x) = \varphi^x \,,
\ee
which is a {\em Fibonacci function} according to the definition of 
Ref.~\cite{HanKimNeggers}, {\em i.e.},
\be
f(x+2)=f(x+1)+f(x) \;\;\;\;\;\; \forall x \in \R \,.
\ee
However, $F_{(e,o)}(x)$ are not Fibonacci functions since they contain 
$\varphi^{-x}$, which is not a Fibonacci function (it is easy to prove \cite{HanKimNeggers}  
that, among power-law functions $y(x)=b^x$, the only Fibonacci function is 
the one with base equal to the golden ratio, $b= \varphi$).

In the following section, we briefly recall the basics of spatially 
homogeneous and isotropic cosmology in general relativity, and then we 
present the formal analogy with Eqs.~(\ref{ODE1})-(\ref{2ndorder}) in 
Sec.~\ref{sec:3}. We then deduce Lagrangian and Hamiltonian formulations 
for the Fibonacci ODEs and derive from these an invariant of the 
(discrete) Fibonacci sequence. 

\section{FLRW cosmology} 
\setcounter{equation}{0}
\label{sec:2}

In the context of Einstein's relativistic theory of gravity, the 
most basic assumption of cosmology is the Copernican principle stating that, 
on average ({\em i.e.}, on scales larger than a 
few tens of Megaparsecs),  
the universe is spatially homogeneous and isotropic 
 \cite{HawkingEllis,Wald,Carroll,KolbTurner,LiddleLyth,Liddle}. 
The stringent symmetry requirements of spatial homogeneity and isotropy force  the geometry to have 
constant spatial curvature \cite{Eisenhart}. The spacetime metric is necessarily given by the 
Friedmann-Lema\^itre-Robertson-Walker (FLRW) line 
element, written in polar comoving coordinates $\left(t,r,\vartheta, \phi 
\right)$ as  
 \cite{HawkingEllis,Wald}
\begin{equation}
ds^2 = -dt^2 + a^2(t) \Bigg[ \frac{dr^2}{1-Kr^2} + r^2 d\Omega_{(2)}^2 
\Bigg]
\end{equation}
where $d\Omega_{(2)}^2 = d\vartheta^2 + \sin^2 \vartheta \, d\phi^2$ is 
the 
line 
element on the unit 2-sphere and the sign of the constant curvature index 
$K$ classifies the 3-dimensional spatial sections $t=$~const. $K>0$ 
corresponds to positively curved 3-spheres, $K=0$ to Euclidean flat 
sections, and $K<0$ to hyperbolic 3-sections. The dynamics of FLRW 
cosmology is contained in the scale factor $a(t)$. In FLRW cosmology, the 
matter source causing the spacetime to curve is usually (but not 
necessarily) taken to be a perfect fluid with energy density $\rho(t)$ and 
pressure $P(t)=w\rho(t)$, where $w$ is a constant equation of state 
parameter. The evolution of $a(t)$ and $ \rho(t)$  is ruled by the 
Einstein equations adapted to the high degree of symmetry, the 
Einstein-Friedmann equations. They comprise  
\cite{HawkingEllis,Wald,Carroll} the 
Friedmann   equation
\begin{equation}
H^2 =   \frac{8 \pi G}{3} 
\rho - \frac{K}{a^2} +\frac{\Lambda}{3} \label{eq:Friedmann}
\end{equation}
(a first order constraint),  the acceleration or Raychaudhuri equation 
\begin{equation}
\frac{\ddot{a}}{a}= -\frac{4 \pi G}{3} \left( \rho + 3 P \right) +  
\frac{\Lambda}{3} \,,\label{eq:acceleration}
\end{equation}
and the covariant conservation equation 
\be
\dot{\rho}+3H \left( P+\rho \right)=0 \label{eq:conservation}
\ee
expressing energy conservation for the cosmic fluid. Here $G$ is Newton's 
gravitational constant, $\Lambda$ is Einstein's 
famous cosmological constant, an overdot 
denotes differentiation with respect to the 
cosmological time $t$,  $H 
\equiv \dot{a}/a$ is the Hubble function \cite{Wald,Carroll}, 
and units 
are used in which the speed of light is unity.\footnote{We follow the 
notations of Refs.~\cite{Wald,Carroll}.} For a generic cosmic 
fluid, only 
two of the three equations~(\ref{eq:Friedmann})-(\ref{eq:conservation}) 
are 
independent. When there is no cosmic fluid and the cosmological constant 
$\Lambda$ (which 
can be treated as an effective fluid with energy density 
$ \rho=\frac{ \Lambda}{8\pi G} =-P$) is the only energy content, the 
conservation equation~(\ref{eq:conservation}) is satisfied identically.

\section{The cosmological analogy}
\setcounter{equation}{0}
\label{sec:3}

Let us consider first the function $F_{(e)}(x)$, which satisfies
\be
F_{(e)}'= \frac{2}{\sqrt{5}} \, \ln\varphi \cosh\left( x\ln\varphi \right)=  
\frac{2}{\sqrt{5}} \, \ln\varphi \sqrt{ 1+\frac{5F_{(e)}^2 }{4} } \,.\label{questa}
\ee
Dividing this equation by $F_{(e)}$ and squaring, one obtains
\be
\left( \frac{F_{(e)}'}{F_{(e)}} \right)^2=  \frac{4\ln^2\varphi}{5} 
+\frac{4\ln^2\varphi}{5F_{(e)}^2} \,,
\ee
which is formally analogous to the Friedmann 
equation~(\ref{eq:Friedmann}), provided that 
$\left(x, F_{(e)}(x) \right) \rightarrow \left(t, a(t) \right)$ and 
\begin{eqnarray}
\rho &=& P= 0 \,,\\
&&\nonumber\\
\Lambda &=& 3\ln^2 \varphi   \,,\\
&&\nonumber\\
K &=& -\frac{4\ln^2\varphi}{5}
\end{eqnarray}
(contrary to FLRW cosmology, these quantities are dimensionless in the 
Fibonacci side of the analogy). 
{\em A priori} the formal equivalence with the Friedmann 
equation~(\ref{eq:Friedmann}) is not 
sufficient for the analogy to hold and one must check that also 
Eqs.~(\ref{eq:acceleration})   
and (\ref{eq:conservation}) are satisfied: this is 
straightforward to do using Eq.~(\ref{2ndorder}), while the conservation 
equation is 
trivially 
satisfied with $\rho=P=0$ (and also by the $\Lambda$-effective fluid with 
$P=-\rho=$~const.). The universe with scale factor analogous to the function $F_{(e)}(x)$ has hyperbolic 
3-D 
spatial sections and is empty, but expands due to the repulsive 
cosmological constant.

By squaring Eq.~(\ref{questa}), one has introduced the possibility of 
solutions with $F'_{(e)}<0$, 
corresponding to a  contracting, instead of expanding, universe. In fact, 
Eq.~(\ref{eq:Friedmann}) then gives
\be
\dot{a} =\pm \sqrt{ \frac{\Lambda a^2}{3} + |K|} \,,
\ee
which integrates to 
\be
\ln \left[ C  \left( \sqrt{ \Lambda^2 a^2 +3\Lambda |K|} +\Lambda a\right)\right] =\pm \sqrt{ 
\frac{\Lambda}{3}} \left( t-t_0 \right) 
\ee
where $C$ and $t_0$ are integration constants ($C$ serves the purpose of making the argument 
of the logarithm dimensionless since, in the units used 
\cite{HawkingEllis,Wald,Carroll},  the scale factor $a$ 
carries the dimensions of a length and $\Lambda$  those of an inverse length squared). By 
exponentiating both sides, squaring, and collecting similar terms, one is left with 
\be
a(t) =a_0 \left[ \mbox{e}^{\pm \sqrt{ \frac{\Lambda}{3}} \left(t-t_0\right)} 
-3|K| \Lambda C^2 \, \mbox{e}^{\mp \sqrt{ \frac{\Lambda}{3}} \left(t-t_0\right)} \right] \,,
\ee
with $a_0$ constant. In practice, in FLRW cosmology the dimensionless 
radial coordinate $r$ can be rescaled to normalize the curvature index $K$ 
(usually to $\pm 1$); here we can rescale to obtain $|K|\Lambda C^2 =1/3$ 
and
\be
a(t)=\pm a_0 \sinh \left[ \sqrt{\frac{\Lambda}{3}}  \left( t-t_0\right) \right] 
\ee
where, in order to keep the scale factor non-negative, the upper sign applies for $t\geq t_0$ and 
the lower one for $t\leq t_0$. At late times $t\rightarrow +\infty$, this metric is asymptotic to 
the metric of de Sitter spacetime $a_{(dS)}(t) = a_0 \, \mbox{e}^{ \sqrt{\frac{\Lambda}{3} } 
\left( t-t_0\right) }$.

Let us focus now on the function $F_{(o)}(x)$: proceeding as we did for $F_{(e)}(x)$, one obtains
\be
\left( \frac{F_{(o)}'}{F_{(o)}} \right)^2= -\ln^2\varphi +\frac{4\ln^2\varphi}{5F_{(o)}^2} \,,
\ee
which is formally analogous to the Friedmann 
equation~(\ref{eq:Friedmann}) with 
\begin{eqnarray}
\rho &=& P= 0 \,,\\
&&\nonumber\\
\Lambda &=& 3\ln^2 \varphi   \,,\\
&&\nonumber\\
K &=& \frac{4\ln^2\varphi}{5} \,.
\end{eqnarray}
The analogous FLRW universe is again empty and propelled by the positive cosmological constant, 
but the spatial sections are now closed 3-spheres. 

Again, by squaring one introduces the possibility of a negative sign for 
$\dot{a}$.  Proceeding in parallel with what has been done for  $F_{(e)}$, 
one obtains
\be
a(t) =a_0 \left[ \mbox{e}^{\pm \sqrt{ \frac{\Lambda}{3}} \left(t-t_0\right)} 
+3K \Lambda C^2 \, \mbox{e}^{\mp \sqrt{ \frac{\Lambda}{3}} \left(t-t_0\right)} \right] 
\ee
and, normalizing $K$ suitably, 
\be
a(t) = a_0 \cosh \left[ \sqrt{\frac{\Lambda}{3}} \left( t-t_0\right) \right] \,.
\ee
This scale factor describes a universe contracting from a infinite size, bouncing at the 
minimum value $a_0$, and then expanding forever and asymptoting to the de Sitter space, a behaviour 
sought for in quantum cosmology to avoid the classical Big Bang singularity  
$a=0$.

Finally, we can consider the Fibonacci function $f(x)= \varphi^x$, which trivially satisfies 
\be
\left( \frac{f'}{f}\right)^2=\ln^2 \varphi 
\ee
and is analogous to an empty ($\rho=P=0$), spatially flat ($K=0$) universe expanding 
exponentially, $a_{(dS)}=a_0 \, \mbox{e}^{\sqrt{\frac{\Lambda}{3}} \, t}$  due to the positive 
cosmological constant $\Lambda=3\ln^2\varphi$. This is the maximally symmetric de 
Sitter universe, which is an attractor in inflationary models of the early universe 
\cite{KolbTurner,LiddleLyth} and in dark energy-dominated models of the late (present-day) 
accelerating universe \cite{AmendolaTsujikawa}.

\section{Lagrangian and Hamiltonian}
\setcounter{equation}{0}
\label{sec:4}

Both the mechanical analogy of Eq.~(\ref{2ndorder}) and the cosmological 
analogy suggest the 
Lagrangian for $F_{(e,o)}(x)$
\be
L_{(e,o)} \left( F_{(e,o)}(x), F_{(e,o)}' (x) \right) =  \frac{1}{2} \, 
(F_{(e,o)}')^2 + 
\frac{\ln^2\varphi}{2} \, F_{(e,o)}^2 \,.
\ee
The corresponding Euler-Lagrange equation
\be
\frac{d}{dx} \left( \frac{ \partial L_{(e,o)} }{ \partial  F_{(e,o)}'  }\right) 
-\frac{\partial L_{(e,o)} }{\partial F_{(e,o)} } =0
\ee
reproduces Eq.~(\ref{2ndorder}). 

The associated Hamiltonian, expressed in terms of the canonical variables $q_{(e,o)} \equiv F_{(e.o)} $ 
and 
$p_{(e,o)}\equiv \partial L_{(e,o)} /\partial F_{(e,o)}' = F_{(e,o)}'$, is
\be
{\mathcal H}_{(e,o)} =p_{(e,o)} F_{(e,o)}' -L_{(e,o)} = \frac{\left( p_{(e,o)} \right)^2 }{2} 
-\frac{\ln^2 \varphi}{2} \, F_{(e,o)}^2 \,,
\ee
and the Hamilton equations are 
\be
q'_{(e,o)} = \frac{\partial {\mathcal H}_{(e,o)} }{\partial p_{(e,o)}  } = 
p_{(e,o)} = 
F_{(e,o)}'= \ln \varphi \, F_{(o,e)} 
\ee
(which reproduces Eqs.~(\ref{ODE1}), (\ref{ODE2})) and
\be
p_{(e,o)}' = -\frac{\partial {\mathcal H}_{(e,o)} }{\partial q_{(e,o)} } =\ln^2 \varphi 
\, F_{(e,o)} \,.
\ee
Since $L_{(e,o)}$ does not depend explicitly on $x$, the Hamiltonian ${\mathcal H}_{(e,o)}$ is 
conserved, 
\be
\frac{\left(p_{(e,o)} \right)^2}{2} 
-\frac{\ln^2 \varphi}{2} \, F_{(e,o)}^2 =E_{(e,o)} \,,
\ee
where the constants $E_{(e,o)}$ have the meaning of energy of the system. Writing them explicitly, we 
have 
\be
E_{(e,o)} = \frac{ \ln^2 \varphi}{2} \left( F_{(o, e)} + F_{(e,o)}\right) \left( F_{(o, e)} - 
F_{(e,o)}   \right) =\pm \frac{2\ln^2\varphi}{5} \,,
\ee
which gives the first integral $ \left( F_{(o,e)}^2 - F_{(e,o)}^2 \right) 
$ on the Fibonacci side of the analogy. In terms of the discrete Fibonacci 
sequence we have, therefore, the\\\\ 
{\em Proposition:}\\
The quantity
\begin{eqnarray}
I & = & \left[  \left(F_{2m}+F_{2m-1}\right)^2 -\left( F_{2m-1}+F_{2m-2} 
\right)^2 \right] \nonumber\\
&&\nonumber\\
&=& F_{2m}^2 + F_{2m-1}^2 -F_{2m-1}^2 -F_{2m-2}^2 +2\left( F_{2m}F_{2m-1} 
-F_{2m-1} F_{2m-2} \right) \nonumber\\
&& \label{eq:E}
\end{eqnarray}
does not depend on $m$ and is an invariant of the Fibonacci sequence.

To the best of our knowledge, this invariant is not related to known 
invariants (for example, those of the Fibonacci convolution sequences 
\cite{Hoggatt}).

\section{Conclusions}
\setcounter{equation}{0}
\label{sec:5}

The functions $F_{(e,o)}$ associated with the continuum generalization of 
the Fibonacci sequence exhibit analogies with certain spatially 
homogeneous and isotropic universes in FLRW cosmology, and also a 
mechanical analogy with an inverted harmonic oscillator, which we have 
presented. With the help of these analogies, it is easy to derive a 
Lagrangian and a Hamiltonian associated with the ODEs satisfied by the 
functions $F_{(e,o)}$. Then, the conservation of the Hamiltonian (or 
energy) yields a quantity $I$, given by Eq.~(\ref{eq:E}) that is constant 
across the Fibonacci sequence, {\em i.e.}, does not depend on the index 
$n$. This invariant may turn out to be useful in view of the many 
applications of the Fibonacci numbers in mathematics and in the natural 
sciences ({\em e.g.}, \cite{A0,A1,A2,A3,A4,A5,A6}).

\begin{acknowledgments} 

We thank two referees for useful comments. This work is supported by the 
Natural Sciences and Engineering Research Council of Canada (Grant No. 
2016-03803 to V.F.) and by Bishop's University.

\end{acknowledgments}



\end{document}